\documentclass[12pt]{article}
\usepackage[dvips]{graphicx}
\def \bsigma{\mbox{\boldmath $\sigma$}}

\def \bM{\mathbf{M}}
\def \bG{\mathbf{G}}
\def \bJ{\mathbf{J}}
\begin{document}
\begin{flushright}
GEF-TH-02-2006
\end{flushright}
\begin{center}

\noindent {\bf {The contribution of strange quarks to the proton
magnetic moment}} \vskip 25 pt G.Dillon and G.Morpurgo
\end{center}
Universit\`a di Genova and Istituto Nazionale di Fisica Nucleare, Sezione
di Genova.
\footnote{e-mail: dillon@ge.infn.it \hskip0.5cm ;\hskip0.5cm
morpurgo@ge.infn.it}

\vskip 15pt \noindent {\bf Abstract.} We show that from the e.m.
magnetic moments of all octet baryons one cannot determine
$\mu_{p}^{s}$ (the strange proton magnetic moment), contrarily to
a result of Jido and Weise. Using the general QCD parametrization
(GP) we clarify why one can, instead, obtain $\mu_{p}^{s}$ from
the Z magnetic moment (or Sachs $G^{Z}_{M}(0)$) of the proton.
This is due to the Trace terms in the GP of the magnetic moments
and specifically to the fact that the coefficient of
$Tr[Q^{\gamma}P^{s}$] cannot be extracted from the octet
($\gamma$) moments ($Q^{\gamma}$'s are the $u,d,s$ electric
charges and $P^{s}$ is the $s$ quark projector). To measure (or
put an upper limit to) $\mu_{p}^{s}$ by the electroweak $e-p$
scattering experiments, one exploits the small terms proportional
to $Tr[Q^{Z}]$ and $Tr[Q^{Z}P^{s}]$ ($Q^{Z}$ are the electroweak
charges of $u,d,s$). We relate $\mu_{p}^{s}$ in the Beck-McKeown
formula to the $Z$ Trace terms. We estimate a 3-gluon exchange
factor reducing the order of magnitude of
$\vert\mu_{p}^{s}\vert$.
 \vskip20pt
 \baselineskip15pt
 \noindent {\bf 1.Introduction}\vskip5pt
\indent  The $s\bar{s}$ contribution $\mu_{p}^{s}$ to the proton
magnetic moment has been the subject, till 5 years ago (compare
\cite{Beck}), of at least 20 calculations producing values (with
both signs) of $\mu_{p}^{s}$ from 0.8 down to 0.003 in units
$\mu_{N}$; a very recent QCD lattice calculation \cite{Leinweber}
 gives $\mu^{s}$= $-0.046\pm0.019$ $\mu_{N}$. The experimental data
- based on the e.m.-weak interference in e-p scattering at low
$q^{2}$ - are at present unable to reach this precision.
\footnote{The last published value from the SAMPLE experiment
(\cite{Sample}) at $q^{2} =0.1 (GeV^{2})$ is :
$G^{s}_{M}(q^{2}=0.1)(p) = 0.37\pm0.20\pm0.26\pm0.07$. For the
results of other continuing experiments
(HAPPEX, GO and MAINZ) compare \cite{HGM}.}\\
\indent In a most recent calculation on the subject \cite{Weise}
it is asserted that $\mu_{p}^{s}$ can be obtained from a group
theoretical $SU_{3}$ analysis of the octet baryon magnetic
moments; the result is (fit 2) $\mu_{p}^{s}= +0.155 \pm 0.022$ or
(with a more complete calculation- fit 3) $\mu_{p}^{s}= +0.161\pm
0.028$ in units $\mu_{N}$. However the method is incorrect. Even a
perfect experimental knowledge of the magnetic moments of all
octet baryons would not allow to deduce -contrarily to what is
implied in Ref.\cite{Weise}- the contribution of the $s\bar{s}$ to
the proton magnetic moment. We consider this here, because the
general QCD parametrization (GP) reveals some interesting aspects.
The main argument, to be developed below, goes as follows: The
treatment in \cite{Weise} ignores the circumstance that the
$Tr[Q^{\gamma}]$ of the electric charge matrix of $u,d,s$ is zero
whereas the $Tr[Q^{Z}]$ of the electro-weak charge $Q^{Z}$ of
$u,d,s$ is different from zero; this explains why, to find
$\mu_{p}^{s}$, one had to embark in the difficult $\gamma$-Z
interference experiments mentioned above.\\
\indent We will proceed as follows. In Sect.2 we recall the basis
of the GP; in Sect.3 we re-discuss the most general QCD expression
of the octet baryon moments; indeed, to extract the contribution
of the $s\bar{s}$ loops in the proton, one must focus on the Trace
terms in the above expression. Note the following : A Trace term
$Tr[Q^{\gamma}P^{s}]$ -that, in principle, might contribute to
$\mu_{p}^{s}$- appears in the GP of the octet baryon magnetic
moments ($P^{s}$ is a projector on the quark s). But (Sect.3),
once the octet magnetic moments of
$p,n,\Lambda,\Sigma^{\pm},\Xi^{0,-}$ and the $\Sigma^{0}\to
\Lambda \gamma$ transition element are parametrized, an identity
relating the coefficients of the various terms precludes to
extract the coefficient $g_{0}$ in front of $Tr[Q^{\gamma}P^{s}]$
from the magnetic moments. Thus $\mu_{p}^{s}$, cannot be related
(via the term $Tr[Q^{\gamma}P^{s}]$), to the experimental values
of the magnetic moments. To determine $\mu_{p}^{s}$, one needs
(Sect.4) the trace terms $Tr[Q^{Z}]$ and $Tr[Q^{Z}P^{s}]$ from the
$Z$ moments. To show this we will re-derive (Sect.4) by the GP the
Beck-McKeown ``key formula" (Eq.(28) of \cite{Beck}) for
the extraction of $\mu_{p}^{s}\equiv G^{s}_{M}(0)$.\\

\vskip20pt \noindent {\bf 2. The general parametrization: Some
elements}\vskip5pt Because the general QCD parametrization (GP)
has been described in detail (\cite{mo89}, \cite{dm96}), we note
here only its basis, needed to clarify the expressions of the
baryon magnetic moments in the following sections. The name
``general QCD parametrization" recalls that the procedure is
derived exactly from the QCD Lagrangian exploiting only a few
general properties. \footnote{Because the script symbols for the
quark fields used in the QCD Lagrangian in \cite{mo89} and in some
other papers may be confusing, standard symbols $u,d,s$ were
adopted from the first paper cited in \cite{dm96} on.}
\\ \indent The GP applies to a variety of QCD
matrix elements or expectation values. \textit{By integrating on
all internal $q\bar{q}$ and gluon lines}, the method parametrizes
exactly such matrix elements. Thus hadron properties -like hadron
masses and magnetic moments- are written exactly as a sum of some
spin-flavor structures each multiplied by a coefficient. Each
structure (term) has a maximum of three indices. The coefficients
of the various structures decrease with increasing complexity of
the structure, giving rise to a hierarchy of the coefficients.
This ''hierarchy" -see Sect.5 for more details- explains why the
non relativistic quark model (NRQM), that keeps only the simplest
additive (one index) terms, works fairly well; More generally the
GP clarifies the relationship of QCD to constituent quark models.\\
\indent To exemplify, we write the  GP magnetic moment of a baryon
$B$:
\begin{equation}
    \langle \psi_{B}|\bM | \psi_{B}\rangle =
    \langle \phi_{B}|V^{\dag} \bM V| \phi_{B}\rangle
    \label{M}
\end{equation}
In Eq.(\ref{M}) $\bM$ is the exact QCD magnetic moment operator,
which, in the baryon rest system, is:
\begin{equation}
     \mathbf{M} = (1/2)\int d^{3}\mbox{\boldmath $ r \, (r\times j(r)$)}
     \label{exM}
\end{equation}
Here
\begin{equation}
    j_{\mu}(x)= e\bar{q}(x)(1/2)[\lambda_{3}+
    (1/\sqrt{3})\lambda_{8}] q(x)
    \label{emc}
\end{equation}
is the e.m. current expressed in terms of the quark $(u,d,s)$
fields $q(x)$; $|\psi_{B}\rangle$ is the exact eigenstate of the
QCD Hamiltonian for $B$ at rest;
 $|\phi_{B}\rangle$ is an auxiliary three body state of $B$,
factorizable as
\begin{equation}
    |\phi_{B}\rangle=|X_{L=0}\cdot W_{B}\rangle
    \label{aux}
\end{equation}
into a space part $X_{L=0}$ with orbital angular momentum zero and
a spin unitary-spin factor $W_{B}$. The unitary transformation $V$
-applied to the auxiliary state $|\phi_{B}\rangle$- transforms the
latter into $|\psi_{B}\rangle$. Integrating on the space
variables, Eq.(\ref{M}) becomes:
\begin{equation}
    \langle \psi_{B}|\bM | \psi_{B}\rangle =
    \langle \phi_{B}|V^{\dag} \bM V| \phi_{B}\rangle=
    \langle W_{B}| \sum_{\nu} g_{\nu}\textbf{G}_{\nu} (s,f) | W_{B}\rangle
    \label{VMV}
\end{equation}
where the $\textbf{G}_{\nu}$'s are operators depending only on the
spin-flavor variables of the three quarks in $\phi_{B}$ and the
$g_{\nu}$'s are a set of parameters. The meaning of $V$ -that
generates $|\psi_{B}\rangle$ from the $|\phi_{B}\rangle$ is
\cite{mo89} that $V$ dresses the auxiliary state with $q\bar{q}$
pairs and gluons; it also introduces configuration mixing (in
other words, $\phi_{B}$ has L=0, but $\psi_{B}$ is a superposition
of states with different L's). The choice of a factorized
structure (space$\times$spin-flavor) of the auxiliary state
$|\phi_{B}\rangle$  (compare Eq.(\ref{aux})) allows the
elimination of the space coordinates in the GP (the second step in
Eq.(\ref{VMV})). A final remark: The GP is, of course, compatible
with chiral theories (compare on this the recent work by Durand
and Ha \cite{dh}) provided that the chiral Lagrangians considered
have the general QCD properties used by the GP. Chiral theories of
baryons incompatible with the correct QCD Lagrangian do not lead
to the GP (see e.g. Sect.VII of the 2nd paper in \cite{dm96}).
\vskip20pt \noindent {\bf3.Can one extract $\mu_{p}^{s}$ from the
magnetic moments of the octet baryons?}\vskip5pt \indent To derive
the Beck-McKeown formula for $\mu_{p}^{s}$ we need only the GP
expressions of the $p$ and $n$ magnetic moments. But the paper by
Jido and Weise \cite{Weise} is based on the magnetic moments of
\textit{all} octet baryons. Thus we must write them below. It is:
\begin{equation}
M_z(B)=\langle W_B \vert \sum_{\nu =0}^7 g_\nu ({\bf G}_\nu)_z
\vert W_B\rangle \equiv \langle W_B \vert \sum_{\nu =1}^7 \tilde
g_\nu ({\bf G}_\nu)_z \vert W_B\rangle
\label{MzB}
\end{equation}
The fact that Eq.(\ref{MzB}) appears as two identical sums, one
with eight, the other with seven terms will be explained in a
moment. First we clarify the notation and use of Eq.(\ref{MzB}).
The symbols are as follows: To first order in flavor breaking (but
-because $P^{s}=[P^{s}]^{n}$ for any integer n- the trace term
remains the same at all orders in flavor breaking), the
$\textbf{G}_{\nu}$'s are:
\begin{equation}
\begin{array}{lcl}
{\bf G}_0=Tr[Q^\gamma P^s]\sum_i \bsigma_i & &
{\bf G}_1=\sum_i Q^\gamma_i \bsigma_i \\ \\
{\bf G}_2=\sum_i Q^\gamma_i P_i^s \bsigma_i & &
{\bf G}_3=\sum_{i\ne k} Q^\gamma_i \bsigma_k \\ \\
{\bf G}_4=\sum_{i\ne k} Q^\gamma_i P_i^s \bsigma_k &&
{\bf G}_5=\sum_{i\ne k} Q^\gamma_k P_i^s \bsigma_i \\ \\
{\bf G}_6=\sum_{i\ne k} Q^\gamma_i P_k^s \bsigma_i && {\bf G}_7
=\sum_{i\ne j\ne k} Q^\gamma_i P_j^s \bsigma_k
\end{array}
\label{Ginu}
\end{equation}
The sums over i,j,k extend to the three quarks in the $W_{B}$
factor of the auxiliary state $\vert\Phi_{B}\rangle$ of the baryon
$B$,
 $Q^{\gamma}_{i}$ is the electric charge of the i-th quark and
$\bsigma_i$ its (Pauli) spin; $P^{s}_{i}$ is a projection operator
(=1 if the i-th quark is strange, =0 if it is $u$ or $d$). From
Eq.(\ref{MzB}) it appears that the magnetic moments are the
expectation values of the $\bG_{\nu}$ in the
spin-flavor states $W_{B}$ of the baryons.\\
\indent The equality of the two sums (with 8 and 7 parameters) in
Eq.(\ref{MzB}) is due to the following identity
(Ref.\cite{dmtrmo99}) relating the $\bG_{\nu}$'s, that holds for
their expectation values in the $\vert W_{B}\rangle
$'s\footnote{The Eq.(\ref{rel}) holds for any traceless $3\times
3$ diagonal matrix $Q$ -not necessarily $Q^{\gamma}$.}
\begin{equation}
\label{rel} {\bf G}_0=-\frac{1}{3} {\bf G}_1+\frac{2}{3} {\bf
G}_2- \frac{5}{6} {\bf G}_3+ \frac{5}{3} {\bf G}_4+\frac{1}{6}{\bf
G}_5+\frac{1}{6}{\bf G}_6+\frac{2}{3}{\bf G}_7
\end{equation}
The magnetic moments of the octet baryons expressed
[Eqs.(\ref{MzB}),(\ref{Ginu})] in terms of the $\tilde g_\nu$ 's
are (the baryon symbol indicates the magnetic moment):
\begin{equation}
\begin{array}{c}
\label{mfg}
p=\tilde g_1\\ \\
n=-(2/3)(\tilde g_1-\tilde g_3)\\ \\
\Lambda =-(1/3)(\tilde g_1-\tilde g_3+\tilde g_2-\tilde g_5)\\ \\
\Sigma^+=\tilde g_1+(1/9)(\tilde g_2-4\tilde g_4-4\tilde
g_5+8\tilde
g_6+8\tilde g_7)\\ \\
\Sigma^-=-(1/3)(\tilde g_1+2\tilde g_3)+(1/9)(\tilde g_2-4\tilde
g_4+2\tilde g_5-4\tilde g_6-
4\tilde g_7)\\ \\
\Xi^0=-(2/3)(\tilde g_1-\tilde g_3)+(1/9)(-4\tilde g_2-2\tilde
g_4+4\tilde g_5-8\tilde g_6+10\tilde
g_7)\\ \\
\Xi^-=-(1/3)(\tilde g_1+2\tilde g_3)+(1/9)(-4\tilde g_2-2\tilde
g_4-8\tilde g_5-2\tilde g_6- 2\tilde g_7)
\end{array}
\end{equation}
and:
\begin{equation}
\label{sigla} \mu(\Sigma \Lambda)=-(1/\sqrt{3})(\tilde g_1-\tilde
g_3+\tilde g_6-\tilde g_7)
\end{equation}
This Okubo (Ref.\cite{okubo}) equation, relating $\mu(\Sigma
\Lambda)$ to the other magnetic moments -correct to first order in
flavor
breaking- is reobtained, of course, in the GP \cite{mo89}.\\
\indent If written with the $g_\nu$'s rather than with the $\tilde
g_\nu$'s, using the equations (\ref{relg}) below - obtained from
the identity (\ref{rel}) - the above formulas (\ref{mfg}) are all
changed in the same way, by the addition of $(-1/3)g_{0}$ to the
r.h.s. of {\em all} expressions (e.g.
$n=-(1/3)g_0-(2/3)(g_1-g_3)$, etc.):
Eq.(\ref{sigla}) stays unchanged. \\
\begin{equation}
\begin{array}{lclcl}
\label{relg} \tilde g_1=g_1-(1/3)g_0 &;&\tilde
g_2=g_2+(2/3)g_0&;&\tilde
g_3=g_3-(5/6)g_0 \\
\tilde g_4=g_4+(5/3)g_0 &;&\tilde g_5=g_5+(1/6)g_0 &;&\tilde
g_6=g_6+(1/6)g_0 \\
\tilde g_7=g_7+(2/3)g_0
\end{array}
\end{equation}
\indent The $\tilde{g}_{\nu}$'s obtained from the magnetic moments
of $p$, $n$, $\Lambda$, $\Sigma^{\pm}$, $\Xi^{-,0}$ are:
\begin{equation}
\begin{array}{lclclcl}
\label{g1} \tilde g_1=2.793 &;&\tilde g_2=-0.934 &;& \tilde
g_3=-0.076 &;&
\tilde g_4=0.438 \\
\tilde g_5=0.097 &;& \tilde g_6=-0.147 &;& \tilde g_7=0.154 &&
\end{array}
\end{equation}
\indent We now come back to the term $\bf
{G}_{0}$=$Tr[QP^s]\sum_i{\bsigma}_{i}$ in Eq.(\ref{MzB}) which is
the only quantity related to $\mu_{p}^{s}$ in the octet magnetic
moments (as we shall see below). From the Eqs.(\ref{mfg}) above it
results that its coefficient $g_{0}$ cannot be determined from the
octet moments. More precisely: The experimental values of the
magnetic moments determine only the seven $\tilde g_{\nu}$'s -see
Eq.(\ref{mfg})- not $g_{0}$.\\ \indent\textit{This fact alone}
shows why the treatment
and results of Ref.\cite{Weise} are not valid.\\
\indent This statement applies to the fit 2 of Ref.\cite{Weise},
where the five terms involved correspond to five of our eight
terms $g_{\nu}$, and also to their fit 3 because the
$g_{0}Tr[Q^{\gamma}P^{s}]$ term discussed above is the only
possible Trace term. Note: A QCD fit to the octet magnetic moments
(correct at least to first order in flavor breaking) requires
seven parameters, after the use of the identity (\ref{rel}). If
(\ref{rel}) is ignored and a fit is performed with a smaller
number of parameters, one may find incorrectly any value for a
misidentified $g_{0}$.\\
\indent To complete the argument we must state why $g_{0}$ would
be the only coefficient of interest for an (impossible) purely
electromagnetic determination of $\mu_{p}^{s}$. The answer is
straightforward: In the result of any QCD field theoretical
calculation of the magnetic moments, Trace terms implying a charge
are associated to $q\bar{q}$ loops; $Tr[Q^{\gamma}P^{s}]$ is the
only Trace term appearing in the $\gamma$ magnetic moments of the
octet baryons and therefore the
only term that might be related to a loop $s\bar{s}$.\\
\indent So far we considered only the magnetic moments, but the
treatment can be easily extended to the $q^{2}$ dependent Sachs
form factors $G_{M}(q^{2})$ of the baryons. As shown in
\cite{dm99PL}, it is sufficient to operate in the Breit frame and
let the coefficients $g_{\nu}$'s depend on $q^{2}$, the square of
the four-momentum transfer. Thus, although here we will refer to
the magnetic
moments $G_{M}(0)$, the same treatment applies to any $G_{M}(q^{2})$.\\
\vskip20pt \noindent {\bf4.The determination of $\mu_{p}^{s}$ from
the electroweak interference.} \vskip5pt \indent The determination
of $\mu_{p}^{s}$ in the $e-p$ scattering experiments (measuring
the parity non conserving amplitude) is discussed in the GP
rewriting the current as:
\begin{equation}
\label{jmu}
 j_{\mu}(x)= e\bar{q}(x)Qq(x)
\end{equation}
where $Q$ has a structure similar for the $\gamma$ or $Z$ exchange
between $e$ and $p$:
\begin{equation}
  Q= c_{0}\lambda_{0} + c_{3}\lambda_{3} + c_{8}\lambda_{8} \equiv
  Q_{u}P^{u}+Q_{d}P^{d}+Q_{s}P^{s}
  \label{Q}
\end{equation}
Here the $c$'s are some (real) coefficients, $\lambda_{0}$ is the
unit matrix Diag[1,1,1]$\equiv$ \textbf{1}; $\lambda_{0},
\lambda_{3},\lambda_{8}$ commute, as required for a GP calculation
in QCD. In the r.h.s. of Eq.(\ref{Q}) $P^{u},P^{d},P^{s}$ are
projectors on the quark fields $u, d, s$. The unit matrix is there
to include the case of $Q^{Z}$, where $Tr[Q^{Z}]\ne 0$\\
\indent For the e.m case one inserts in Eq.(\ref{jmu})
$Q=Q^{\gamma} = (1/2)[\lambda_{3}+(1/\sqrt{3})\lambda_{8}]$, that
is the diagonal matrix Diag[$Q_{u}^{\gamma}, Q_{d}^{\gamma},
Q_{s}^{\gamma}]\equiv [2/3, -1/3, -1/3]$. It is $Tr[Q^{\gamma}]=0$
(we limit to the quarks $u,d,s$).\\ \indent For the exchange of a
$Z$ the electro-weak charges $Q^{Z}$ of the quarks in Eq.(\ref{Q})
are:
\begin{eqnarray}
Q^{Z}_{u}& = & k_{W}[1-(8/3)\sin^{2}\theta_{W}]\nonumber\\
Q^{Z}_{d} & = & k_{W}[-1+(4/3)\sin^{2}\theta_{W}]\label{QZ}\\
Q^{Z}_{s} & = & k_{W}[-1+(4/3)\sin^{2}\theta_{W}]\nonumber
\end{eqnarray}
where $k_{W}= [2\sin(2\theta_{W})]^{-1}$; that is, $Q^{Z}$ in
Eq.(\ref{jmu}) is the matrix Diag[$Q^{Z}_{u}, Q^{Z}_{d},
Q^{Z}_{s}$]; both $Tr[Q^{Z}]$ and $Tr[Q^{Z}P^{s}]$ do not vanish.
 \\ \indent Now we will
obtain by the GP $\mu^{Z}_{p}$  and the basic formula (Eq.(28) of
\cite{Beck}) giving $\mu^{s}_{p}$ in the electro-weak interference
experiments. \textit{We will show that such formula works due the
fact that $Tr[Q^{Z}]$ and $Tr[Q^{Z}P^{s}]$ do not vanish; the sum
of their coefficients $g_{0}+\hat{g}_{0}$ is what the
experiments are trying to measure}.\\
\indent The current (\ref{jmu}), similar for $\gamma$
and $Z$ exchange, makes the GP procedure simple.\\
\indent From now on we consider only $p$ and $n$. From
Eq.(\ref{VMV})
their magnetic moments are:\\
\begin{equation}
\mathbf{M}^{Q}(p,n)=g_{1}\sum_i Q_i \bsigma_i + g_{3}\sum_{i\ne
k}Q_i \bsigma_k + g_{0}Tr[QP^s]\sum_i \bsigma_i
 + \hat{g}_{0}Tr[Q]\sum_i\bsigma_i\\
 \label{MQpn}
\end{equation}
In Eq.(\ref{MQpn}) $Q$ is the charge given in Eq.(\ref{Q}); to
reproduce the $\gamma$ magnetic moments of $p,n$-(Sect.3)- one
inserts for $Q$ the electric charge diagonal matrix
$Diag[2/3,-1/3,-1/3]$. To obtain the magnetic moments measured in
the parity non conserving electro-weak $e-p$ scattering, one
inserts in Eq.(\ref{MQpn}) the electro-weak charge $Q(Z)$
[Eq.(\ref{QZ})].\footnote{The trace term
$\hat{g}_{0}Tr[Q]\sum_i\bsigma_i$ that now appears in the
Eqs.(\ref{MQpn}), was not present in the $\gamma$ determination of
the magnetic moments, because $Tr[Q^{\gamma}]$ vanishes if only
the $u,d,s$ quarks are considered. [We also saw that
$Tr[Q^{\gamma}P^{s}]$
 does not play a role for the e.m. moments]}\\
\indent We write separately the contributions of
$u,d,s$ to $\mathbf{M}^{Q}$ (for $p, n$); from Eq.(\ref{MQpn}) it is:\\
\begin{equation}
\mathbf{M}^{Q} = \mathbf{M}^{u} + \mathbf{M}^{d} + \mathbf{M}^{s}
\label{sepQ}
\end {equation}
where the $\mathbf{M}^{u}, \mathbf{M}^{d}, \mathbf{M}^{s}$ are
written below [from Eq.(\ref{MQpn})] in terms of the projectors
$P^{u}, P^{d}, P^{s}$:
\begin{equation}
\begin{array}{l}
\mathbf{M}^{u}= g_{1}\sum_i Q_{i}P_i^u \bsigma_i + g_{3}\sum_{i\ne
k}Q_{i}P_i^u
\bsigma_k + \hat{g}_{0}Tr[QP^{u}]\sum_i\bsigma_i. \\ \label{MQflav}\\
\mathbf{M}^{d}=g_{1}\sum_i Q_{i}P_i^d \bsigma_i + g_{3}\sum_{i\ne
k}Q_{i}P_i^d
\bsigma_k +\hat{g}_{0}Tr[QP^{d}]\sum_i\bsigma_i. \nonumber\\ \\
\mathbf{M}^{s}=(\hat{g}_{0} + g_{0})Tr[QP^{s}]\sum_i\bsigma_i
\nonumber
\end{array}
\end{equation}
These equations hold both for the $\gamma$ and $Z$ contributions
to $\mathbf{M}^{u}, \mathbf{M}^{d}, \mathbf{M}^{s}$; that is in
Eqs.(\ref{MQflav}) any $Q$ can be either $Q^{\gamma}$ or $Q^{Z}$.
Note that $\sum_i\bsigma_i = 2\bJ$; of course $2J_{z}=1$ if the baryon spin is up along the z axis.\\
\indent The contributions of the quarks $u, d, s$ to the ($\gamma$
or $Z$) magnetic moment of the proton are the expectation values
$\mu^{f}_{p} = \langle p\uparrow \vert$M$^{f}_{z} \vert p\uparrow
\rangle$
where the upper $f$ in $M^{f}_{z}$ stays for $u$ or $d$ or $s$. Using the Eq.(\ref{MQflav})
and writing also:\\
\begin{equation}
\tilde{g}_{0}\equiv g_{0} + \hat{g}_{0}
\label{g0t}
\end{equation}
one obtains:\\
\begin{equation}
\begin{array}{l}
(\mu^{u}_{p}/Q_{u}) = \frac{4}{3} g_{1} + \frac{2}{3} g_{3}
+\hat{g}_{0}
 \equiv  \frac{4}{3}\tilde{g}_{1} + \frac{2}{3}\tilde{g}_{3} +
\tilde{g}_{0}\\
\label{mupflav}
(\mu^{d}_{p}/Q_{d}) = -\frac{1}{3} g_{1} +
\frac{4}{3} g_{3} + \hat{g}_{0}
 \equiv -\frac{1}{3} \tilde{g}_{1} + \frac{4}{3}\tilde{g}_{3}
+ \tilde{g}_{0} \nonumber \\
(\mu^{s}_{p}/Q_{s}) = \tilde{g}_{0} \nonumber \\
\end{array}
\end{equation}
\\
Similarly for the neutron we have:\\
\begin{equation}
\begin{array}{l}
(\mu^{u}_{n}/Q_{u}) = -\frac{1}{3} g_{1} + \frac{4}{3} g_{3} +
\hat{g}_{0} \equiv -\frac{1}{3} \tilde{g}_{1} +
\frac{4}{3}\tilde{g}_{3} + \tilde{g}_{0}\\
(\mu^{d}_{n}/Q_{d}) = \frac{4}{3} g_{1} + \frac{2}{3} g_{3} +
\hat{g}_{0} \equiv\frac{4}{3}\tilde{g}_{1} +
\frac{2}{3}\tilde{g}_{3} + \tilde{g}_{0}
\label{munflav} \\
(\mu^{s}_{n}/Q_{s}) = \tilde{g}_{0} \nonumber \\
\end{array}
\end{equation}\\
\indent The Eqs.(\ref{mupflav}),(\ref{munflav}) refer to the
$\gamma$ magnetic moments (on inserting there $Q = Q^{\gamma}$) or
to the $Z$ magnetic moments (if $Q = Q^{Z}$). Our
$(\mu^{u}_{p}/Q_{u})$ (and the similar quantities for $d,s$) are
the Sachs form factors [at $q^{2}$=0] $G_{M}^{u}(0)$,
$G_{M}^{d}(0)$, $G_{M}^{s}(0)$ of
Ref.\cite{Beck}.\footnote{Compare their Eqs.(26),(27) that define
$G^{\gamma}_{M}$ and $G^{Z}_{M}$}\\
\indent In the Eqs.(\ref{mupflav},\ref{munflav}) the right hand
side of the expressions
for $[\mu^{u}_{p}/Q_{u}]$ , $[\mu^{d}_{p}/Q_{d}]$, \\
$[\mu^{u}_{n}/Q_{u}]$, $[\mu^{d}_{n}/Q_{d}]$ have been written in
terms of the tilded parameters $\tilde{g}_{1},\tilde{g}_{3}$ using
the Eqs.(\ref{relg}). However recall that the tilded parameters
extracted from the $\gamma$ magnetic moments, give no information
on $g_{0}$ nor, of course, on $\hat{g}_{0}$. Thus the
interpretation of the Eqs.(\ref{mupflav}) and (\ref{munflav}) is
the following: The $s\bar{s}$ contributions to the magnetic
moments, which can only be read in the Trace terms, must be
extracted measuring $\tilde{g}_{0}$ by the very difficult $Z$
exchange experiments (as stated $Tr[Q^{Z}]$ and $Tr[Q^{Z}P^{s}]$
do not vanish). Clearly, from the
Eqs.(\ref{mupflav},\ref{munflav})(multiplying each by its
$Q_{u},\: Q_{d},\: Q_{s}$ and summing) it emerges that, to
determine $\tilde{g}_{0}$, $Tr(Q) \equiv (Q_{u}+Q_{d}+Q_{s})$ has
to be different from
zero.\\
\indent From the Eqs.(\ref{mupflav}) and (\ref{munflav}) the
$\gamma$ and $Z$ moments of $p$ and $n$ follow immediately. They
are (the $\gamma$ moments coincide of course with those of
Eq.(\ref{mfg})):
\begin{equation}
\mu^{\gamma}_{p} = \tilde{g}_{1} \qquad \qquad\mu^{\gamma}_{n} =
-(2/3)(\tilde{g}_{1} - \tilde{g}_{3})
\label{mugamma}
\end{equation}
\begin{equation}
\begin{array}{l}
\mu^{Z}_{p} = k_{W}\Big[((5/3) - 4\sin^{2}\theta_{W})\tilde{g}_{1}
-(2/3)\tilde{g}_{3}
-\tilde{g}_{0}\Big] \label{muzeta}\\
\mu^{Z}_{n} = k_{W}\Big[(-(5/3) + (8/3)
\sin^{2}\theta_{W})\tilde{g}_{1} +((2/3)
-(8/3) \sin^{2}\theta_{W})\tilde{g}_{3} -\tilde{g}_{0}\Big] \nonumber\\
\end{array}
\end{equation}
The first Eq.(\ref{muzeta}) is identical to the Eq.(28) of
\cite{Beck} (transcribed below):
\begin{equation}
G^{Z,p}_{M}= (1-4\sin^{2}\theta_{W})G^{\gamma,p}_{M} -
G_{M}^{\gamma,n} -G^{s}_{M}
\end{equation}
as can be seen recalling that, in the notation of \cite{Beck},
$\mu^{Z}_{p}/k_{W}\equiv G^{Z,p}_{M}$, $\mu_{p}^{\gamma}\equiv
G_{M}^{\gamma,p}$, $\mu_{n}^{\gamma}\equiv G_{M}^{\gamma,n}$,
$\tilde{g}_{0}\equiv G_{M}^{s}$.\\
\indent The novel feature of this GP derivation of the
Beck-McKeown equation is that $\mu^{s}_{p}$ is related directly to
the Trace terms in
a QCD plus electroweak field theoretical description.\\
\vskip20pt \noindent {\bf5. A GP estimate of the 3-gluon exchange
factor reducing $\mu_{p}^{s}$} \vskip5pt \indent The GP hierarchy
of the coefficients means (Sect.2) that the coefficient of each
sum of terms (characterizing the hadron property under study)
decreases when the number of different indices in the terms under
consideration increases or when -at equal number of indices-
flavor breaking factors are present. To exemplify, each
$\bf{G}_{\nu}$'s of the baryon magnetic moments in Sect.3 is a sum
of terms. Other examples, among the many ones, are the octet and
decuplet baryon masses \cite{mo89,dm96} or the electromagnetic
mass differences of baryons \cite{mo92dmmir}. The order of
magnitude of the reduction -mentioned above- of the coefficients
is $\approx 0.33$ for the reduction due to a flavor breaking
factor and $0.33\pm 0.05$ for an index due to the \textit{exchange
of one gluon between two ``constituent" quark lines}.
\begin{figure}
    \begin {center}\includegraphics[width=7cm,height=6cm]{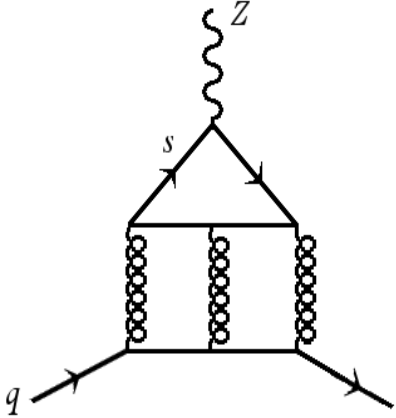}
   \end{center}
   \caption{\footnotesize{The $Z$ interacts with a $s$-quark 
loop (in the $e-p$ parity violating scattering) that must be connected by at least three 
gluons to an internal quark line $q$ of the proton. Note: The three gluons may also reach 
different internal quark lines, not the same as in the figure.}}
   \label{1}
   \end{figure}
Here, to estimate $\mu^{s}_{p}$, we are interested in the
reduction factor due to the three gluons forming a colour singlet
state with J=1. If, for each gluon in Fig.1, the value of the
reduction factor were the same ($\approx 1/3$) as that for gluon
exchange between constituent quarks inside a hadron, one would
have, approximately, a reduction factor of the order $(1/3)^{3}$;
however, for this case (Fig.1), an independent estimate is appropriate.\\
\indent This estimate is possible applying the GP analysis
\cite{mVP} of the vector meson $\gamma$ decays to the $\phi \to
\pi\gamma$ and $\omega \to \pi\gamma$. One sees from the table I
of \cite{mVP}) that the rate of the $\phi\to\pi\gamma$ is
determined by the deviation $\Delta\theta_{V}$ of the vector
mixing angle $\Delta\theta_{V}$ from its ideal value ($35.3^{o}$)
plus two terms $\sqrt{\frac{2}{9}}\cdot\Gamma_{5}$ and
$\sqrt{\frac{2}{3}}\cdot\Gamma_{7}$ produced by a three gluon
(J=1) exchange between $\phi$ and $\pi^{0}$. Although
$\Delta\theta_{V}$ is itself due to a three gluon $J=1$ color
singlet exchange, below we will focuss on these terms
$\Gamma_{5}$ and $\Gamma_{7}$.\\
\indent The uncertainty arising from the poor knowledge of the
form factor in the $\phi\to\pi\gamma$ decay can be partially
eliminated by considering the ratio between the $\phi\to\pi\gamma$
and the $\omega\to\pi\gamma$ decay rates, because the two decay
momenta (379 and 501 Mev) are not too different and the
$\omega\to\pi\gamma$ rate is reasonably well known. The $\rho
\to\pi\gamma$ is presently affected by much larger errors. From
the table I of \cite{mVP} cited above and writing an equation
similar to the Eq.(68) of \cite{mVP} (but now with the numerator
referring to $\phi$ and the denominator to $\omega$) we have
\footnote{In the table I of \cite{mVP} all terms proportional to
$\sin\Delta\theta_{V}$ were ignored except for that of the direct
$\phi\to \pi^{0}\gamma$ decay.}:
\begin{equation}
\frac{\Gamma(\phi\to\pi^{0}\gamma)}{\Gamma(\omega\to\pi^{0}\gamma)}=
\Bigg\vert\frac{\sin\Delta\theta_{V}\cdot\mu_{1}\:f_{1}(k_{\phi})+
\sqrt{\frac{2}{9}}\:\mu_{5}\:f_{5}(k_{\phi})+\sqrt{\frac{2}{3}}\:\mu_{7}\:
f_{7}(k_{\phi})}{\mu_{1}\:f_{1}(k_{\omega}) +
\frac{2}{3}\:\mu_{5}\:f_{5}(k_{\omega})}\Bigg\vert^{2}\cdot
\frac{k_{\phi}^{3}}{k_{\omega}^{3}} \label{ratio}
\end{equation}
Note \cite{mVP} that the two terms proportional to $f_{5}$ and
$f_{7}$ in the numerator of Eq.(\ref{ratio}) correspond to a three
gluon transition of an ideal $\phi$ (a pure $s\bar{s}$ state) into
an ideal $\omega$ followed by the $\pi_{0}+\gamma$ decay of the
latter. The ratio $x$ (to be introduced below) is therefore
related to the transition of an $s\bar{s}$ state into a
$u\bar{u}+d\bar{d}$ state both with $J=1$; it gives some estimate
(though in a different situation) of the 3-gluon exchange factor
reducing $\mu^{s}_{p}$. We
now proceed with the calculation of the r.h.s. of Eq.(\ref{ratio}). \\
\indent In Eq.(\ref{ratio}) the term $\mu_{5}$ in the denominator
can be neglected. Assuming equal form factors for $\phi$ and
$\omega$ and defining $x$:
\begin{equation}
               x\equiv
               \textstyle\Big[\sqrt{\frac{2}{9}}\:\mu_{5}+
               \sqrt{\frac{2}{3}}\:\mu_{7}\Big]\Big/\mu_{1}
\end{equation}
we obtain:
\begin{equation}
\frac{\Gamma(\phi\to\pi^{0}\gamma)}{\Gamma(\omega\to\pi^{0}\gamma)}=
2.3\:\Big\vert \sin\Delta\theta_{V} + x \Big\vert^{2}\
\label{sratio}
\end{equation}
where the factor $2.3$ is the ratio $[k_{\phi}/k_{\omega}]^{3}$.
Because
$\Big[{\Gamma(\phi\to\pi^{0}\gamma)}\Big/(2.3)\cdot{\Gamma(\omega\to\pi^{0}\gamma)}\Big]=(3.4\pm
0.2)\cdot10^{-3}$ it is:
\begin{equation}
\Big\vert \sin\Delta\theta_{V} + x \Big\vert^{2} = (3.4\pm
0.2)\cdot10^{-3} \label{sqmod}
\end{equation}
Setting $\sin\Delta\theta_{V}=0$ (in the linear description -
implied by the GP- $\sin\Delta\theta_{V}$ is small $\approx
1.7\cdot 10^{-2}$)-, the right hand side of Eq.(\ref{sqmod}) gives
for the three gluon reduction factor $\vert x\vert $ of
interest\footnote{On the other hand a value of
$\sin\Delta\theta_{V}$ corresponding to quadratic mixing
($\theta_{V}=38.7^{0}$) might give rise to a vanishingly small
$\vert x \vert $.}:
\begin{equation}
 \vert x\vert \approx 5.8\cdot 10^{-2}
 \label{xval}
\end{equation}
Note that $\vert x\vert$ is expressed in units of the $p$ magnetic
moment ($2.79\:\mu_{N}$). To estimate the correct order of
magnitude of $\vert\mu^{s}_{p}\vert$ one should know, however, in
addition to $\vert x\vert $, two quantities: a) The factor arising
from the various ways of connecting the three gluons from the loop
to the quark lines in the proton; b) The factor arising because
the three gluons originate from a $s\bar{s}$ loop, different from
the situation intervening in the above calculation of $x$. We are
unable to calculate these factors but we felt that the strong
reduction $\vert x\vert $ might be of some interest in itself. To
give an idea of the number in Eq.(\ref{xval}), note that if the
product of the factors a),b) above produces a further reduction by
2 or 3, we are in the region of values of $\vert\mu^{s}_{p}\vert$
given in \cite{Leinweber}; a region where the experiments are
-perhaps optimistically- called in \cite{Leinweber} ``tremendously
challenging".
\pagebreak

\end{document}